\documentclass[11pt,a4paper]{article}

\usepackage[T1]{fontenc}
\usepackage[utf8]{inputenc}
\usepackage{lmodern}
\usepackage{microtype}
\usepackage[a4paper,margin=1in]{geometry}
\usepackage{setspace}
\usepackage{lineno}
\usepackage[numbers,sort&compress]{natbib}

\usepackage{amsmath,amssymb,amsthm}
\newcommand{\argmin}{\mathop{\rm arg~min}\limits}
\newtheorem{lemma}{Lemma}
\newtheorem{theorem}{Theorem}
\newtheorem{prop}{Proposition}
\newtheorem{rmk}{Remark}

\usepackage{graphicx}
\usepackage{tikz}
\usepackage[export]{adjustbox}
\usepackage{subcaption}
\usepackage{float}
\usepackage{booktabs}
\usepackage{multirow}
\usepackage[boxruled,linesnumbered]{algorithm2e}

\usepackage[hidelinks]{hyperref}

\usepackage{authblk}

\newif\ifanonymous
\anonymousfalse

\title{Optimal Post-processing of Synthetic Data for Pearson Correlation Matching}

\ifanonymous
  \author{Anonymous Author(s)}
  \date{}
\else
  \author[1]{Oussama Ounissi\thanks{Corresponding author: \href{mailto:ounissioussama@stu.kanazawa-u.ac.jp}{ounissioussama@stu.kanazawa-u.ac.jp}}}
  \author[2]{Nicklas J\"averg\r{a}rd}
  \author[3]{Assaad Zeghina}
  \author[2]{Adrian Muntean}

  \affil[1]{Graduate School of Natural Science and Technology, Kanazawa University, Kanazawa 920-1192, Ishikawa, Japan}
  \affil[2]{Department of Mathematics and Computer Science, Karlstad University, Karlstad 651 88, Sweden}
  \affil[3]{LATMOS/IPSL, UVSQ Université Paris-Saclay, Sorbonne Université, CNRS., Guyancourt 78280, France}
  \date{}
\fi

\begin{document}

\maketitle

\ifanonymous\else
\begin{center}
\small
\href{mailto:nicklas.javergard@kau.se}{nicklas.javergard@kau.se}\quad
\href{mailto:assaad.zeghina@latmos.ipsl.fr}{assaad.zeghina@latmos.ipsl.fr}\quad
\href{mailto:adrian.muntean@kau.se}{adrian.muntean@kau.se}

\vspace{0.35em}
ORCID: \href{https://orcid.org/0009-0008-4962-358X}{0009-0008-4962-358X} (O. Ounissi); 
\href{https://orcid.org/0000-0002-9185-4209}{0000-0002-9185-4209} (N. J\"averg\r{a}rd); 
\href{https://orcid.org/0000-0002-2649-5609}{0000-0002-2649-5609} (A. Zeghina); 
\href{https://orcid.org/0000-0002-1160-0007}{0000-0002-1160-0007} (A. Muntean)
\end{center}
\fi

\begin{abstract}
Preserving correlation in synthetic data is of interest in several applications. Existing approaches mainly address correlation preservation during the generation procedure. Here, we instead consider it as a postprocessing problem. Given original and synthetic tabular data, we seek the smallest change to the synthetic dataset that restores the Pearson correlation matrix of the original data. Under suitable assumptions, we derive a unique explicit solution to the minimization problem. We also provide a bound on the size of the correction in terms of the initial correlation error. On the numerical side, across several datasets and generation methods, the proposed approach largely preserves marginal distributions, t-SNE geometry, and classification performance. The method is generator-independent and can be applied after synthesis without modifying the generation procedure.
\end{abstract}


\vspace{0.75em}

\section{Introduction}

\label{sec1}

Synthetic data generation has attracted considerable attention in recent years due to its potential applications in privacy preservation, data sharing, and data scarcity, such as in the case of class imbalance. A wide variety of approaches have been proposed in the literature, ranging from statistical and sampling-based methods to machine learning techniques relying on generative neural networks~\citep{FonsecaBacao2023,adasyn,sdv,ctgan}.
One important aspect in the usefulness of synthetic data is its ability to preserve the dependence structure of the original dataset. In particular, correlation preservation is relevant in applications such as synthetic health data~\citep{ZhangMuller2026} and power-system data~\citep{ChatzosEtAl2022}. Consequently, several works have investigated the generation of synthetic data while preserving correlation-related quantities~\citep{Ruscio2008, YANG2022, Nicklas_SD}.

Most of these approaches address dependence preservation during the generation procedure. A complementary possibility is to correct an already generated synthetic dataset. This can be useful when the generation procedure is fixed or otherwise satisfactory, but a particular statistical property needs to be corrected. Compared with the extensive literature on synthetic data generation, less attention has been given to such postprocessing approaches. In this direction, \citet{WangEtAl2023} proposed a postprocessing approach based on resampling techniques, with correlation alignment among the considered applications.

This motivates the following problem: given an already generated synthetic dataset, can its correlation structure be restored while changing the data as little as possible? In this work, we address this question by proposing a lightweight postprocessing approach for synthetic tabular data. Starting from a given synthetic dataset, obtained through an arbitrary generation procedure, we construct the closest dataset whose Pearson correlation structure matches that of the original data while preserving the mean and standard deviation of the synthetic data. In that sense, the proposed method should be viewed as a correction or enhancement step that can be applied on top of existing synthetic data generators.

On the theoretical side, we first establish a characterization of feature sets having the same Pearson correlation structure through the action of linear orthogonal maps in the centered-data subspace. This naturally leads to an orthogonal minimization problem. However, it is necessary to ensure that the resulting correction remains in the centered-data subspace, which does not follow immediately from the classical Orthogonal Procrustes problem. We therefore prove a result ensuring this property under suitable assumptions, which also guarantees the uniqueness of the corrected data. We also provide a bound relating the size of the correction to the error between the starting correlation matrices and present an efficient implementation for the case where the number of features is relatively small.
A closely related approach is ADICOV~\citep{CamachoEtAl2011}, which considers least-squares approximation of a dataset under a prescribed covariance structure and employs the Orthogonal Procrustes problem in its construction. While the resulting constructions are closely related, the present work provides the necessary rigorous justification of the reduction to the Orthogonal Procrustes problem in the Pearson-correlation setting.

Applications to several datasets and synthetic data generators illustrate the effectiveness of the proposed approach. In particular, the numerical experiments indicate that the correlation structure can be restored while largely preserving the individual feature distributions, the t-SNE geometry of the data, and the performance of downstream classification tasks.

\section{Background}\label{back}
In this section, we introduce the setting, the notation, and a classical result used throughout the paper.
Given a set of features (original data), arranged in a matrix $O = [O_1,\dots, O_m] \in \mathbb{R}^{n\times m}$, $m \leq n$, we denote by $S = [S_1,\dots,S_{m}] \in \mathbb{R}^{p\times m}$ a synthetic dataset that shares certain statistical characteristics with $O$.
We implicitly assume that no feature is a constant number for the Pearson correlation to make sense. We also assume that $n \leq p$. This assumption is natural in the context of synthetic data. Moreover, it is necessary to establish the characterization given in Lemmas \ref{lemcos} and \ref{lempc}. In particular, the converse implication does not hold without it.
Furthermore, we assume that $O$, and occasionally other data matrices, are of full rank, \textit{i.e.} admit a left inverse denoted by ${}^\dagger$. This assumption is motivated mathematically by the fact that the set of left-invertible matrices has full measure in $\mathbb{R}^{n\times m}$ for $m \leq n$. In the case of $O$, it ensures the existence of a linear map $SO^{\dagger}$ satisfying $(SO^{\dagger})O = S$.

Let $\|\cdot\|$ and $\langle \cdot, \cdot \rangle$ denote the Frobenius norm and inner product. We denote $I$ as the identity matrix and $\mathbf{1} = [1,\dots,1]^T$, with the dimension understood from the context. We note that when referring to a non-square orthogonal matrix, it is understood that either its rows or columns are orthonormal. 
For a feature $f = [f^1, \cdots, f^n]^T \in \mathbb{R}^n$, we define its arithmetic mean by $\operatorname{Mean}(f) = \frac{1}{n}\sum_{i=1}^n f^i$. The mean centering of $f$ is denoted by $\bar f = [f^1 - \operatorname{Mean}(f), \cdots, f^n-\operatorname{Mean}(f)]^T$, and its standard deviation by $\operatorname{Std}(f) = \frac{1}{\sqrt{n}}\|\bar f\|$.
For a matrix $A = [A_1, \dots,A_m] \in \mathbb{R}^{n\times m}$, we denote by $\operatorname{Col}(A)$ and $\operatorname{Row}(A)$ the vector subspaces generated by its columns and rows, respectively.
When $A$ is symmetric positive-definite, 
$A^{1/2}$ denotes its unique symmetric positive-definite square root. The inverse of $A^{1/2}$ is denoted by $A^{-1/2}=(A^{1/2})^{-1}$.
We define the mean and the standard deviation vectors of $A$ as $\operatorname{Mean} (A) = [\operatorname{Mean} (A_i)]_{i=1,\dots m} \in \mathbb{R}^{m}$ and $\operatorname{Std} (A) = [\operatorname{Std} (A_i)]_{i=1,\dots m} \in \mathbb{R}^{m}$, respectively.
We also define the cosine similarity and Pearson correlation matrices of $A$, respectively, by
\[
\operatorname{S_c}(A) =
\left[
\frac{\langle A_i, A_j \rangle}{\|A_i\|\|A_j\|}
\right]_{i,j=1,\dots,m}
\in \mathbb{R}^{m\times m},
\quad
\operatorname{Corr}(A) =
\left[
\frac{\langle \bar A_i, \bar A_j \rangle}{\|\bar A_i\|\|\bar A_j\|}
\right]_{i,j=1,\dots,m}
\in \mathbb{R}^{m\times m}.
\]

In the following proposition, we introduce the classical Orthogonal Procrustes problem (see \cite{Bisgard}) .

\begin{prop}[Orthogonal Procrustes problem] \label{OPP}
Let $A,B \in \mathbb{R}^{n\times m}$. Then,
\[
Q_*= \argmin_{QQ^T=I}\|QA-B\| = U V^T,
\]
where $U,V^T \in \mathbb{R}^{n\times n}$ are orthogonal matrices and $\Sigma \in \mathbb{R}^{n\times n}$ is a diagonal matrix with non-negative entries obtained from the Singular Value Decomposition (SVD) of
\[
BA^T = U \Sigma V^T.
\]
\end{prop}
We note that generally $Q_*$ is not unique.

\section{Methodology}\label{result}

First, we give the following lemma, which plays a crucial role in the proof of Theorem~\ref{main-result}. In particular, it guarantees that the Orthogonal Procrustes step remains in the centered-data subspace, thereby justifying the use of the standard Orthogonal Procrustes problem in our setting. It also directly yields the uniqueness of the solution in Theorem~\ref{main-result} and underlies the efficiency improvement noted in Remark~\ref{remcomp}.

\begin{lemma}\label{OPPsub}
Let $A,B \in \mathbb{R}^{n\times m}$, $m\leq n$, with $A$ and $B$ of full rank. Set
\[
Q_* = \argmin_{QQ^T=I}\|QA-B\|.
\]
Then it follows that  $\operatorname{Col}(Q_* A) \subset \operatorname{Col}(B)$. Moreover, $Q_* A$ is unique.
\end{lemma}

\begin{proof}
From Proposition \ref{OPP}, $Q_*$ is given by $Q_* = U V^T$ with $BA^T = U \Sigma V^T$.

Since $A$ and $B$ are of full rank, we have  $\operatorname{rank}(BA^T) = m$. We consider the reduced SVD
\[
BA^T = U_m \Sigma_m V^T_m,
\]
where $U_m \in \mathbb{R}^{n\times m}, V^T_m \in \mathbb{R}^{m\times n}$ are orthogonal matrices and $\Sigma_m \in \mathbb{R}^{m\times m}$ is invertible. We note that 
\begin{align}
    \Sigma_m^{-1}U_m^TBA^T = V^T_m,\label{sv} \\
    BA^T V_m\Sigma_m^{-1} = U_m.\label{su}
\end{align}
From Eq.~\eqref{sv} follows $\operatorname{Row}(V_m^T) = \operatorname{Row}(A^T)$. Therefore, we have
\[
  U_m V_m^T A = U V^T A = Q_* A.
\]
From Eq.~\eqref{su} follows $\operatorname{Col}(U_m) = \operatorname{Col}(B)$. Therefore, we have 
\[
 \operatorname{Col}(U_m V_m^T A) \subset \operatorname{Col}(B).
\]
The uniqueness of $Q_*A$ follows since $U_m V_m^T$ is unique.    
\end{proof}
Under the aforementioned assumption, we establish complete characterizations of feature sets having identical cosine similarity and Pearson correlation matrices. 
These characterizations are the key ingredient in the explicit construction of synthetic data derived in Theorem \ref{main-result}.
We start by presenting in Lemma \ref{lemcos} a necessary and sufficient condition for two feature sets to have the same cosine similarity matrix, which represents the cosine of the angle between every two features.

\begin{lemma}\label{lemcos}
Let $O \in \mathbb{R}^{n\times m}$ and $S \in \mathbb{R}^{p\times m}$, with $m\le n \le p$, and let $O$ be of full rank. Then,
\[
\operatorname{S_c}(O) = \operatorname{S_c}(S)
\]
if and only if there exist an orthogonal matrix $M \in \mathbb{R}^{p\times n}$ and a diagonal matrix $N \in \mathbb{R}^{m\times m}$ with strictly positive entries, such that
\[
M O N = S.
\]
\end{lemma}

\begin{proof}
Extend $O$ by zero rows such that $n=p$, and by orthonormal unit vectors $O_i \in \langle O_1,\ldots,O_{i-1} \rangle^\perp$ for $i>m$, and similarly for $S$; such that $m=n=p$.
Let $O, S \in \mathbb{R}^{p\times p}$, satisfying $\operatorname{S_c}(O) = \operatorname{S_c}(S)$. Take $N =\operatorname{diag}\Bigl(\frac{\|S_1\|}{\|O_1\|},\cdots, \frac{\|S_p\|}{\|O_p\|} \Bigr)$, and set $O_N = ON$. We have $O_N^TO_N = S^TS$, and from the setting, $O_N$ admits an inverse $O_N^{-1}$. Taking $M = S O_N^{-1} $, we have $M O N = S$ and
\begin{align*}
    M^TM &= (O_N^{-1})^T S^T S  O_N^{-1} = I.
\end{align*}

For the converse implication, let $O \in \mathbb{R}^{n\times m}, S \in \mathbb{R}^{p\times m}$, such that $MON = S$, with $N = \operatorname{diag}\Bigl(N_1,\cdots, N_m \Bigr) $ and $M \in \mathbb{R}^{p\times n}$ is orthogonal. Consider an appropriate extension of $O$ and $M$, as previously noted, such that $n=p$. 
Then, for $1\le i,j \le m $, we have
\begin{align*}
    \frac{\langle S_i, S_j \rangle}{\|S_i\|\|S_j\|} 
    &= \frac{\langle N_iMO_i, N_jMO_j \rangle}{\|N_iMO_i\|\|N_jMO_j\|} \\
    & =  \frac{\langle O_i, M^TMO_j \rangle}{\|O_i\|\|O_j\|} = \frac{\langle O_i, O_j \rangle}{\|O_i\|\|O_j\|}.
\end{align*}

The assertions are completed by removing the appropriate rows/columns if necessary.
\end{proof}
 
The condition \(n\le p\) is necessary for the converse implication. For example, set
\[
M =
\begin{bmatrix}
1 & 0 & 0\\
0 & 1 & 0
\end{bmatrix}
\in \mathbb{R}^{2\times 3},
\quad
O =
\begin{bmatrix}
0 & 1\\
1 & 2\\
2 & 3
\end{bmatrix}
\in \mathbb{R}^{3\times 2}, \quad
S =
\begin{bmatrix}
0 & 1\\
1 & 2
\end{bmatrix}
\in \mathbb{R}^{2\times 2}.
\]
We have $MO = S$, however $\operatorname{S_c}(O) \neq \operatorname{S_c}(S)$.

Now we address the case of Pearson's correlation. Let
\[
D := \frac{1}{n}\mathbf{1}\mathbf{1}^T \in \mathbb{R}^{n\times n}.
\] 
The set of features with zero mean (centered-data), denoted $\mathcal{C}_n = \{f \in \mathbb{R}^n;\ \operatorname{Mean}(f) = 0\} = \langle \mathbf{1} \rangle^\perp$, is a hyperplane subspace of $\mathbb{R}^n$ perpendicular to $\mathbf{1}$.
Mean centering a feature $f \in \mathbb{R}^n$ is given by $\bar{f} = (I-D)f$, which is the orthogonal projection onto the subspace $\mathcal{C}_n$. For $A\in \mathbb{R}^{n\times m}$, we define 
\[
\bar{A} := (I-D)A,
\]
and note that $\operatorname{Col}(A) \subset \mathcal{C}_n$ if and only if $\bar A = A$.

The Pearson correlation between $f,g \in \mathbb{R}^n$ is the cosine similarity between their respective orthogonal projections onto $\mathcal{C}_n$. Pearson's correlation coincides with the cosine similarity whenever $f, g \in \mathcal{C}_n$. In particular,
\begin{equation}\label{percos}
    \operatorname{Corr}(A) = \operatorname{S_c}(\bar A).
\end{equation}

Parallel to Lemma \ref{lemcos}, we present in Lemma \ref{lempc} a necessary and sufficient condition for two feature sets to have the same Pearson correlation matrix.

\begin{lemma} \label{lempc}
Let $O \in \mathbb{R}^{n\times m}$ and $S \in \mathbb{R}^{p\times m}$, with $m< n \le p$, and let $\bar O$ be of full rank. Then,
\[
\operatorname{Corr}(O) = \operatorname{Corr}(S)
\]
if and only if there exist an orthogonal matrix $M \in \mathbb{R}^{p\times n}$ and a diagonal matrix $N \in \mathbb{R}^{m\times m}$ with strictly positive entries, such that
\[
M\bar{O}N = \bar{S}.
\]
\end{lemma}

\begin{proof}
The conclusion follows immediately from \eqref{percos} and Lemma \ref{lemcos}.     
\end{proof}

For given $O$ and $S$, now we are in a position to answer the question of the closest matrix $\hat S$ to $S$ having the same Pearson correlation matrix as $O$. Thanks to the previous characterization, this question boils down to finding the matrices $M$ and $N$. The matrix $N$ controls the standard deviation of the resulting $\hat S$, and thus can be fixed first to achieve a desired standard deviation. In the context of synthetic data, it is often chosen to be the same as that of $S$. With a fixed $N$, finding $M$ is then achieved through the Orthogonal Procrustes problem. All of this takes place in the subspace of centered-data, and is made possible by Lemma \ref{OPPsub}. Finally, by translation, we obtain $\hat S$ with a desired mean, which in this context is often chosen to be the same as that of $S$.

\begin{theorem}\label{main-result} 
Let $O \in \mathbb{R}^{p\times m}$ and $S \in \mathbb{R}^{p\times m}$, with $m < p$, and let $\bar O$ and $\bar S$ be of full rank. The unique closest matrix $\hat S$ to $S$ in the Frobenius norm, with given $\operatorname{Mean}(\hat S) = [\mu_1, \dots, \mu_m]$ and $\operatorname{Std}(\hat S) = [\sigma_1, \dots, \sigma_m]$, $\sigma_i >0$, for all $1\le i \le m$; such that $\operatorname{Corr}(\hat S) = \operatorname{Corr}(O)$, is given by
\begin{align}
    \hat{S} = M \bar{O}N + T, \quad M = U V^T;
    \label{eq:svd-construct}
\end{align}
where $U , V^T \in \mathbb{R}^{p\times p}$ are orthogonal matrices, and $\Sigma \in \mathbb{R}^{p\times p}$ is a diagonal matrix with non-negative entries obtained from the SVD of
\[
\bar{S}(\bar{O}N)^T = U \Sigma V^T.
\]
With
\[
T = \mathbf{1} \mu^T,
\quad \mu = [\mu_1,\dots,\mu_m]^T \in \mathbb{R}^m;
\qquad
N =\operatorname{diag}\Bigl(\frac{\sigma_1}{\operatorname{Std}(\bar O_1)},\cdots, \frac{\sigma_m}{\operatorname{Std}(\bar O_m)} \Bigr).
\]
\end{theorem}

\begin{proof} 
We note that in Proposition \ref{OPP}, if $A, B$ are of full rank and $\operatorname{Col}{B} \subset \mathcal{C}_p$, then by Lemma \ref{OPPsub} we have
\[
\argmin_{\substack{QQ^T = I\\ \operatorname{Col}(QA) \subset \mathcal{C}_p}} \|QA-B\|
=
\argmin_{QQ^T = I} \|QA-B\|.
\]

From Lemma \ref{lempc}, we deduce that $\hat{S}$ is of the form $\hat{S} = M\bar{O}N + T$, where $T$ and $N$ are defined as in the statement of the theorem. In this context, $T$ and $N$ are uniquely determined by the desired $\operatorname{Mean}(\hat S)$ and $\operatorname{Std}(\hat S)$, respectively. Finally, we obtain $M$ by letting $A:= \bar{O} N$ and $B:= \bar{S}$ in Proposition \ref{OPP}. The uniqueness of $\hat S$ follows from Lemma \ref{OPPsub}.
\end{proof}

\begin{rmk}
The previous result applies more generally to $O \in \mathbb{R}^{n\times m}$, for $n \le p$. It suffices to extend $\bar O$ by zero rows and adjust the normalization accordingly.
\end{rmk}

The constraint on the standard deviation is naturally motivated from the point of view of synthetic data. From a mathematical point of view, this plays to our advantage, since to make use of the closed form of $M$ provided by the Orthogonal Procrustes problem, the matrix $N$ has to be fixed a priori. However, fixing $N$ is not merely done for mathematical convenience, as we show in the following remark.
\begin{rmk}
If $N$ is not fixed a priori, the minimization problem may fail to admit a solution, as in the following counterexample.
Consider $O,S\in\mathbb{R}^{3\times2}$, such that $\bar O$ and $\bar S$ have unit columns, with $\angle(\bar O_1,\bar O_2)=\pi/6$ and $\angle(\bar S_1,\bar S_2)=5\pi/6$.
Let $N=\operatorname{diag}(N_1,N_2)$, with $N_1,N_2>0$, and assume that the minimum
\[
l
=
\min_{\substack{MM^T=I\\N_1,N_2>0}}
\|M\bar O N-\bar S\|
\]
is attained. From Figure~\ref{fig:N_counterexample}, it is not difficult to show that $l>1$. Choose an orthogonal $\widetilde M$ such that $\widetilde M\bar O_1=\bar S_1$,
and set $N_\varepsilon=\operatorname{diag}(1,\varepsilon)$, $\varepsilon>0$. Then
\[
\|\widetilde M\bar O N_\varepsilon-\bar S\|\longrightarrow 1
\qquad\text{as }\varepsilon\to0^+.
\]
Hence, for $\varepsilon$ sufficiently small, $\|\widetilde M\bar O N_\varepsilon-\bar S\|<l$, a contradiction. Therefore, the minimum is not attained.
\end{rmk}

\begin{figure}[H]
\centering
\begin{tikzpicture}[scale=2.4, >=stealth]

\tikzset{
    Ovec/.style={->, very thick, blue},
    Svec/.style={->, very thick, red}
}

\node[anchor=west] at (-1.15,1.15) {$\mathcal C_3$};

% S vectors
\draw[Svec] (0,0) -- (1.2,0)
    node[right, red] {$\bar S_1$};

\draw[Svec] (0,0) -- ({1.2*cos(150)},{1.2*sin(150)})
    node[above left, red] {$\bar S_2$};

% O vectors
\draw[Ovec] (0,0) -- ({1.2*cos(60)},{1.2*sin(60)})
    node[above right, blue] {$\bar O_1$};

\draw[Ovec] (0,0) -- ({1.2*cos(90)},{1.2*sin(90)})
    node[above, blue] {$\bar O_2$};

% angle between O columns
\draw[thin, blue]
    ({0.42*cos(60)},{0.42*sin(60)})
    arc[start angle=60,end angle=90,radius=0.42];
\node[blue] at ({0.54*cos(75)},{0.54*sin(75)})
    {$\frac{\pi}{6}$};

% angle between S columns
\draw[thin, red]
    (0.68,0)
    arc[start angle=0,end angle=150,radius=0.68];
\node[red] at ({0.82*cos(25)},{0.82*sin(25)})
    {$\frac{5\pi}{6}$};

\end{tikzpicture}

\caption{The column verctors of $\bar O$ and $\bar S$ in the centered-data subspace $\mathcal C_3$.}
\label{fig:N_counterexample}
\end{figure}

Prescribing the mean is only done for generality. Since the correlation depends only on the centered data, the mean can be prescribed independently by translation. Moreover, if the mean is not prescribed and the objective is simply to find the closest admissible matrix to $S$, then the optimal choice is $\operatorname{Mean}(\hat S) =\operatorname{Mean}(S)$. We therefore use the mean and standard deviation of $S$ as the natural choices in Algorithm~\ref{alg:ppcorr}.

For the natural choice of the prescribed mean and standard deviation proposed in Algorithm \ref{alg:ppcorr}, we give in the following proposition, a bound for the correction error in terms of the error in the correlation matrices and the scale of the starting synthetic data $S$.
\begin{prop}
Let $O$, $S$, $\hat S$ be as in Theorem~1. Set
$\operatorname{Mean}(\hat S)=\operatorname{Mean}(S)$,
$\operatorname{Std}(\hat S)=\operatorname{Std}(S)$,
and let $\Lambda = \operatorname{diag} \bigl( \operatorname{Std}(S_1), \dots, \operatorname{Std}(S_m)\bigr)$, $\sigma_{\max} = \max_{1\leq i\leq m} \operatorname{Std}(S_i)$.
Then
\[
\|\hat S-S\|
\leq
\sqrt{p}\,\sigma_{\max}
\left\|
\operatorname{Corr}(O)^{1/2}
-
\operatorname{Corr}(S)^{1/2}
\right\|.
\]
\end{prop}

\begin{proof}
Since $\bar S$ and $\bar O$ are of full rank and $\Lambda$ 
is invertible, $\operatorname{Corr}(S)$ and
$\operatorname{Corr}(O)$ are symmetric positive definite. 
Set
\[
Q=
\frac{1}{\sqrt{p}}\,
\bar S\Lambda^{-1}\operatorname{Corr}(S)^{-1/2}.
\]
We note that 
\[
\frac{1}{p}\bar S^T\bar S = \Lambda\operatorname{Corr}(S)\Lambda,
\]
then we have
\begin{equation} \label{Qorth}
Q^TQ = \operatorname{Corr}(S)^{-1/2}
\Lambda^{-1} \left(\frac{1}{p}\bar S^T\bar S\right)
\Lambda^{-1} \operatorname{Corr}(S)^{-1/2} =I
\end{equation}
and
\[
\bar S
=
\sqrt{p}\,Q\operatorname{Corr}(S)^{1/2}\Lambda.
\]
Consider
\[
Y=
\sqrt{p}\,Q\operatorname{Corr}(O)^{1/2}\Lambda.
\]
It is easy to check that $\bar Y = Y$, $\operatorname{Std}(Y) = \operatorname{Std} (S)$, and $\operatorname{Corr}(Y) = \operatorname{Corr} (O)$, then by Theorem \ref{main-result} and \eqref{Qorth} we have
\begin{align*}
\|\hat S-S\| & \leq \| Y-\bar S\|\\
&= \sqrt{p}\, \left\| \left( \operatorname{Corr}(O)^{1/2}
- \operatorname{Corr}(S)^{1/2} \right)\Lambda \right\|\\
&\leq \sqrt{p}\,\sigma_{\max} \left\| \operatorname{Corr}(O)^{1/2} - \operatorname{Corr}(S)^{1/2} \right\|.
\end{align*}
\end{proof}
In particular, by Theorem~6.2 in \cite{Higham2008}, we have
\[
\|\operatorname{Corr}(O)^{1/2}-\operatorname{Corr}(S)^{1/2}\|
\leq
\frac{\|\operatorname{Corr}(O)-\operatorname{Corr}(S)\|}
{\sqrt{\lambda_{\min}(\operatorname{Corr}(O))}+
\sqrt{\lambda_{\min}(\operatorname{Corr}(S))}},
\]
where $\lambda_{\min} (\cdot)$ gives the smallest eigenvalue. Thus, away from nearly singular correlation matrices, a small difference in correlation requires a small correction of the synthetic data.

Lemma~\ref{OPPsub} also leads to the computational optimization detailed in the following remark.
\begin{rmk}\label{remcomp}
By Lemma \ref{OPPsub}, we only need to calculate the reduced SVD of $\bar{S} (\bar{O}N)^T$. If $m\ll p$, this can be done efficiently via reduced QR decompositions:
\[
\bar{S}=Q_S R_S,
\qquad
\bar{O}N = Q_O R_O,
\]
where $Q_S,Q_O\in\mathbb{R}^{p\times m}$ are orthogonal and $R_S,R_O\in\mathbb{R}^{m\times m}$. Then
\[
\bar{S} (\bar{O}N)^T
=
Q_S (R_S R_O^T) Q_O^T.
\]
Thus, by the SVD of $R_S R_O^T=\widetilde U\widetilde \Sigma\widetilde V^T$, we obtain
\[
\bar{S} (\bar{O}N)^T
=
(Q_S\widetilde U)\widetilde\Sigma(Q_O\widetilde V)^T.
\]
\end{rmk}

The post-processing procedure, with the mean and the standard deviation chosen to match those of $S$, is summarized in Algorithm 1.
\begin{algorithm}
\DontPrintSemicolon
\caption{Procrustes-based Pearson Correlation Post-processing}
\label{alg:ppcorr}
\KwIn{$O \in \mathbb{R}^{n \times m}$, $S \in \mathbb{R}^{p \times m}$, with $m < n \le p$.}
\KwOut{$\hat{S}$ with $\operatorname{Corr}(\hat{S}) = \operatorname{Corr}(O)$, $\operatorname{Std} (\hat{S}) =\operatorname{Std} (S)$, $\operatorname{Mean} (\hat{S}) =\operatorname{Mean} (S)$.}
\BlankLine
\SetKwFunction{FPP}{ppcorr}
\SetKwFunction{FExt}{data\_ext}
\SetKwFunction{FSVD}{SVD}
\SetKwProg{Fn}{Function}{:}{end}
\Fn{\FPP{$O, S$}}{
    $\bar{O} \gets O - \mathbf{1} \operatorname{Mean}(O)^T$     \tcp*[r]{center original}
    $\bar O \gets \FExt( \bar O, p)$                                       \tcp*[r]{zero-pad if $n < p$}
    $\bar{S} \gets S - \mathbf{1} \operatorname{Mean}(S)^T$     \tcp*[r]{center synthetic}
    $A \gets \bar{O}\, \operatorname{diag}\Bigl(\frac{\operatorname{Std}(\bar S_1)}{\operatorname{Std}(\bar O_1)},\cdots, \frac{\operatorname{Std}(\bar S_m)}{\operatorname{Std}(\bar O_m)} \Bigr)$                                       \tcp*[r]{}
    $U, \Sigma, V^T \gets \FSVD(\bar{S} A^T)$                    \tcp*[r]{}
    $M \gets U\, V^T$                                \tcp*[r]{}
    $\hat{S} \gets M A + \mathbf{1} \operatorname{Mean}(S)^T$ \tcp*[r]{cf.~Eq.~\eqref{eq:svd-construct}}
    \KwRet $\hat{S}$
}
\end{algorithm}

\section{Experimental Evaluation} \label{exper}

Theorem~\ref{main-result} guarantees that $\hat{S}$ achieves exact Pearson correlation matching with $O$, but leaves open the question of how the transformation affects marginal feature distributions, geometry, and downstream utility. This section investigates these aspects empirically across diverse datasets and generation methods.

\subsection{Datasets and Experimental Setup}
\label{sec:setup}

\paragraph{Datasets.}
We evaluate our approach on five benchmark datasets covering diverse domains and scales. The Credit Card Fraud dataset~\citep{creditcard} contains 284{,}807 transactions with 28 PCA-derived numerical features and severe class imbalance. The MAGIC Gamma Telescope dataset~\citep{magic} provides 19{,}020 simulated particle shower observations described by 10 continuous features, with a binary signal-versus-background target. The Covertype dataset~\cite{covtype} consists of 581{,}012 forest patches described by 10 cartographic features across 7 cover type classes. The Dry Bean dataset~\citep{drybean} contains 13{,}611 seed samples with 16 morphological features across 7 bean varieties. Finally, SustData dataset~\cite{Pereira2014} contains electricity production and consumption measurements from households in Madeira Island, comprising 5 million one-minute averaged observations across five features. The dataset carries no class labels.

\paragraph{Generation methods.}
For each dataset, generating synthetic samples up to a fixed target size using five generation methods: SMOTE~\citep{smote}, ADASYN~\citep{adasyn}, Gaussian Copula, TVAE~\citep{sdv}, and CTGAN\citep{ctgan}. This selection covers the main families of synthetic data generation: nearest-neighbor interpolation, density-adaptive oversampling, parametric modelling, and deep generative approaches. Our Orthogonal Procrustes post-processing is then applied to the generated samples. 
For clarity of presentation, $O$ denotes the original dataset and $S$ denotes a dataset generated by some method; 
We denote by $\hat{S}$ the resulting dataset from applying the transformation indicated in Theorem.~\ref{main-result}. 
We seek from $\hat{S}$ to preserve the same mean and standard deviation vectors of $S$. 

Furthermore, $\hat{O}$ denotes the resulting dataset from applying Theorem~\ref{main-result} on the original dataset $O$.
It is applied exclusively on the SustData dataset to indicate that the approach works as intended, \textit{i.e.}, the closest data to $O$ with an identical feature correlation is $O$ itself.

% ________________________________________________________________
\subsection{Numerical Validation of Correlation Matching}
\label{sec:corr_results}

We quantify the effect of our post-processing step through three metrics comparing the Pearson correlation matrix of the synthetic data $C_S$ to that of the original training data $C_O$, before and after correction: the \textbf{Mean Absolute Correlation Difference} $\mathrm{MACD} = \frac{1}{m^2}\sum_{i,j}|C_O^{ij} - C_S^{ij}|$, the \textbf{Maximum Deviation} $\mathrm{MaxDev} = \max_{i,j}|C_O^{ij} - C_S^{ij}|$, and the \textbf{Frobenius Distance} $\|C_O - C_S\|_F$.

\begin{table}[!htbp]
\centering
\caption{Correlation metrics before (B) and after (A) Orthogonal Procrustes post-processing, averaged across all five generation methods. All metrics are $\downarrow$ (lower is better). After correction all values are at numerical zero.}
\label{tab:corr_metrics}

\begin{tabular}{lcccccc}
\toprule
 & \multicolumn{2}{c}{\textbf{MACD} $\downarrow$} 
 & \multicolumn{2}{c}{\textbf{MaxDev} $\downarrow$} 
 & \multicolumn{2}{c}{\textbf{FrobCorr} $\downarrow$} \\
\cmidrule(lr){2-3}\cmidrule(lr){4-5}\cmidrule(lr){6-7}
\textbf{Dataset} & $S$ & $\hat{S}$ & $S$ & $\hat{S}$ & $S$ & $\hat{S}$ \\
\midrule
Credit Card     & 0.1148 & $  0$ & 0.4699 & $  0$ & 4.8297  & $  0$ \\
MAGIC           & 0.0709 & $  0$ & 0.2853 & $  0$ & 1.1506  & $  0$ \\
Covertype       & 0.0627 & $  0$ & 0.2134 & $  0$ & 0.8383  & $  0$ \\
Dry Bean        & 0.1097 & $  0$ & 0.3399 & $  0$ & 2.3523  & $  0$ \\
SustData        & 0.2503 & $  0$ & 0.9935 & $  0$ & 1.8093  & $  0$ \\
\bottomrule
\end{tabular}
\end{table}

\begin{figure}
\centering
\begin{adjustbox}{scale=0.67}
\begin{tikzpicture}{scale=0.8}
  % First row
  \node[anchor=north west] (img1) at (0,0) 
      {\includegraphics[width=.32\textwidth,height=.32\textwidth]{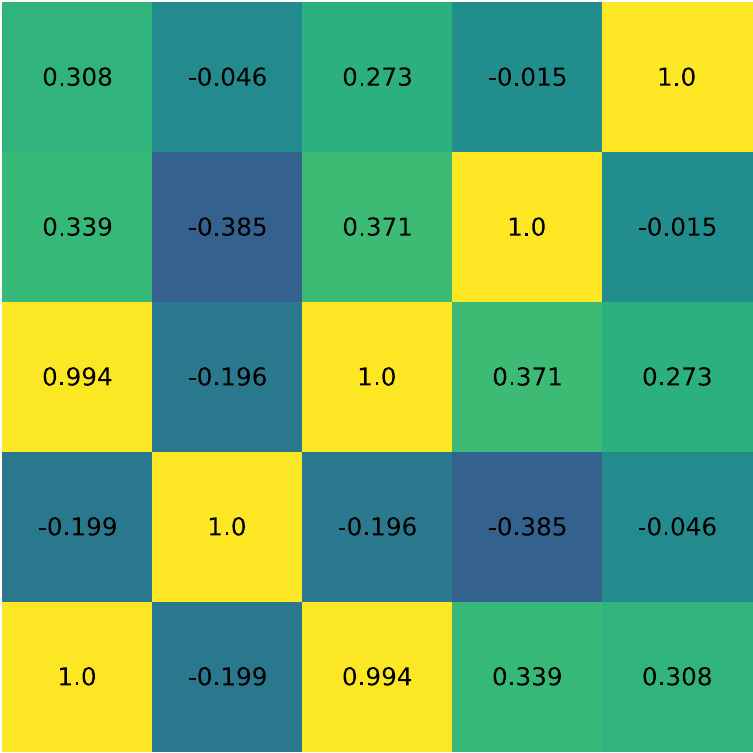}};
  \node[anchor=north west] (img2) at (6,0) 
      {\includegraphics[width=.32\textwidth,height=.32\textwidth]{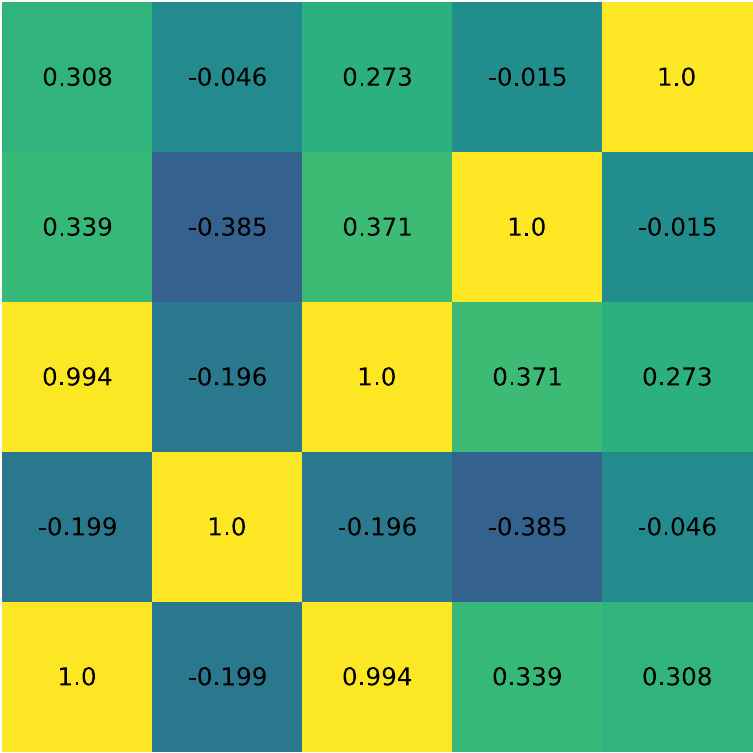}};
  
  % Second row
  \node[anchor=north west] (img3) at (0,-7) 
      {\includegraphics[width=.32\textwidth,height=.32\textwidth]{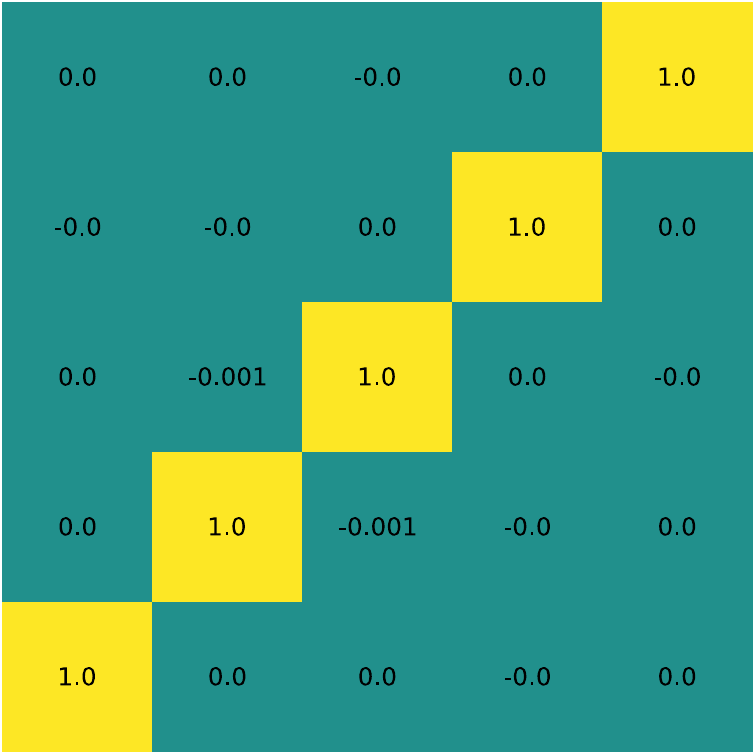}};
  \node[anchor=north west] (img4) at (6,-7) 
      {\includegraphics[width=.32\textwidth,height=.32\textwidth]{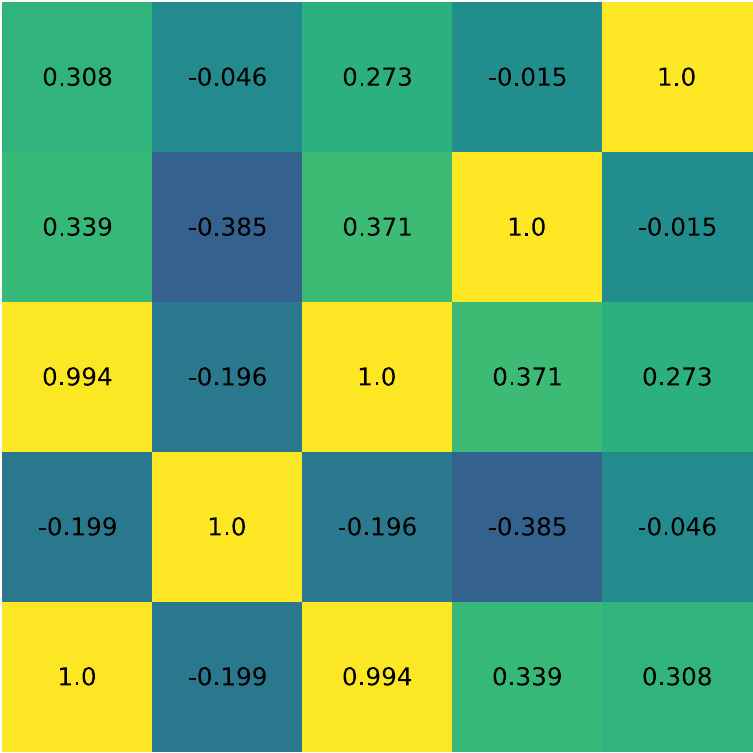}};
  \node at (2.9,0.3) {$O$};
  \node at (2.9+6,0.3) {$\hat{O}$};
  \node at (2.9,0.3-7) {$S$};
  \node at (2.65+6,0.3-7) {$\hat{S}$};
  % ---- Example annotations ----
  % Add text labels
  \node at (0.7,-5.8) {$v_1$};
  \node at (1.8,-5.8) {$v_2$};
  \node at (2.8,-5.8) {$v_3$};
  \node at (3.9,-5.8) {$v_4$};
  \node at (5,-5.8) {$v_5$};

  \node at (0.7+6,-5.8) {$v_1$};
  \node at (1.8+6,-5.8) {$v_2$};
  \node at (2.8+6,-5.8) {$v_3$};
  \node at (3.9+6,-5.8) {$v_4$};
  \node at (5+6,-5.8) {$v_5$};

  \node at (-0.2,-5.0) {$v_1$};
  \node at (-0.2,-3.95) {$v_2$};
  \node at (-0.2,-2.9) {$v_3$};
  \node at (-0.2,-1.85) {$v_4$};
  \node at (-0.2,-0.80) {$v_5$};

  \node at (0.7,-5.8-7) {$v_1$};
  \node at (1.8,-5.8-7) {$v_2$};
  \node at (2.8,-5.8-7) {$v_3$};
  \node at (3.9,-5.8-7) {$v_4$};
  \node at (5,-5.8-7) {$v_5$};

  \node at (0.7+6,-5.8-7) {$v_1$};
  \node at (1.8+6,-5.8-7) {$v_2$};
  \node at (2.8+6,-5.8-7) {$v_3$};
  \node at (3.9+6,-5.8-7) {$v_4$};
  \node at (5+6,-5.8-7) {$v_5$};

  \node at (-0.2,-5.0-7) {$v1$};
  \node at (-0.2,-3.95-7) {$v2$};
  \node at (-0.2,-2.9-7) {$v3$};
  \node at (-0.2,-1.85-7) {$v4$};
  \node at (-0.2,-0.80-7) {$v5$};
  
\end{tikzpicture}
\end{adjustbox}
\caption{Pearson correlation matrices. From top to bottom and left to right: $O$, $\hat O$, $S$, and $\hat S$.}
\label{fig:Correlations}
\end{figure}

Table~\ref{tab:corr_metrics} reports the three correlation metrics averaged across all five generation methods per dataset. The results are consistent: across all datasets and methods, the post-processing step reduces every metric to machine precision, with Frobenius distances in the range $[10^{-15}, 10^{-13}]$, constituting a numerical confirmation of the theoretical guarantee of Theorem~\ref{main-result}. The magnitude of the initial correlation distortion varies across generation methods and datasets, yet in all cases the post-processing step reduces every metric to machine precision, demonstrating that the approach is equally effective regardless of the generator used. Figure~\ref{fig:Correlations} provides a visual illustration of this result on the SustData dataset, where $S$ is generated by independently sampling each feature's empirical marginal distribution, deliberately discarding all inter-feature structure. As expected, applied to $O$, the method does not interfere at all with the data \textit{i.e.}, $\hat O = O$. We also see how it realigns the correlations of $S$ such that the correlations of $\hat{S}$ are identical to $O$.

\subsection{Geometric Analysis via t-SNE}
\label{sec:tsne}
To complement the quantitative analysis, we examine the geometric effect of the post-processing step using two-dimensional t-SNE~\citep{tsne} embeddings. For each generation method, we jointly embed the real minority training samples $O$, the raw synthetic samples $S$, and the post-processed samples $\hat{S}$ into a shared low-dimensional space. This unified embedding allows direct comparison of the relative positions of the three distributions within the same coordinate system.

\begin{figure*}
    \centering
    \includegraphics[width=\linewidth]{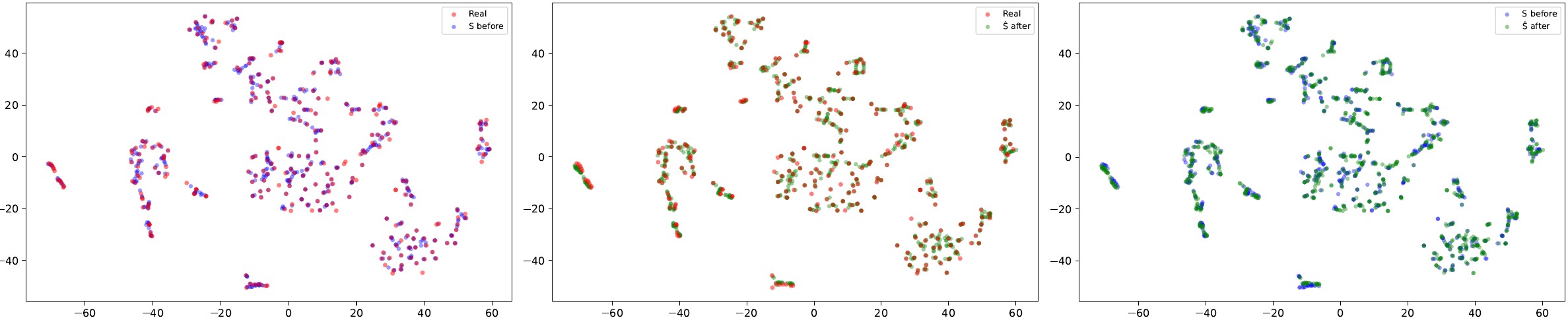}
    \caption{t-SNE visualization on Credit Card (CTGAN). Left: real $O$ (red) vs.\ raw synthetic $S$ (blue). Center: real $O$ (red) vs.\ post-processed $\hat{S}$ (green). Right: $S$ (blue) vs.\ $\hat{S}$ (green).}
    \label{fig:tsne_example}
\end{figure*}

Figure~\ref{fig:tsne_example} illustrates the joint t-SNE embedding for the Credit Card dataset generated with CTGAN. The real samples $O$ remain fixed across the left and center panels, serving as a reference, while the right panel compares $S$ and $\hat{S}$ directly.
The key observation is not whether $\hat{S}$ resembles $O$ more closely, but rather whether the post-processing distorts the geometric structure of $S$ itself. Comparing the left and right panels, the post-processed samples $\hat{S}$ retain the overall point cloud geometry of $S$ without collapsing or significantly reorganizing its structure, indicating that enforcing the correlation constraint does not substantially alter the distributional character of the synthetic data. The transformation can therefore be understood as a structured realignment of the inter-feature relationships within $S$, rather than a resampling or redistribution of its points.
Similar patterns are observed across all datasets and generation methods: the post-processing step modifies the correlation structure of $S$ while leaving its overall geometric footprint largely intact.

\begin{figure*}
    \centering
    \includegraphics[width=0.9\linewidth]{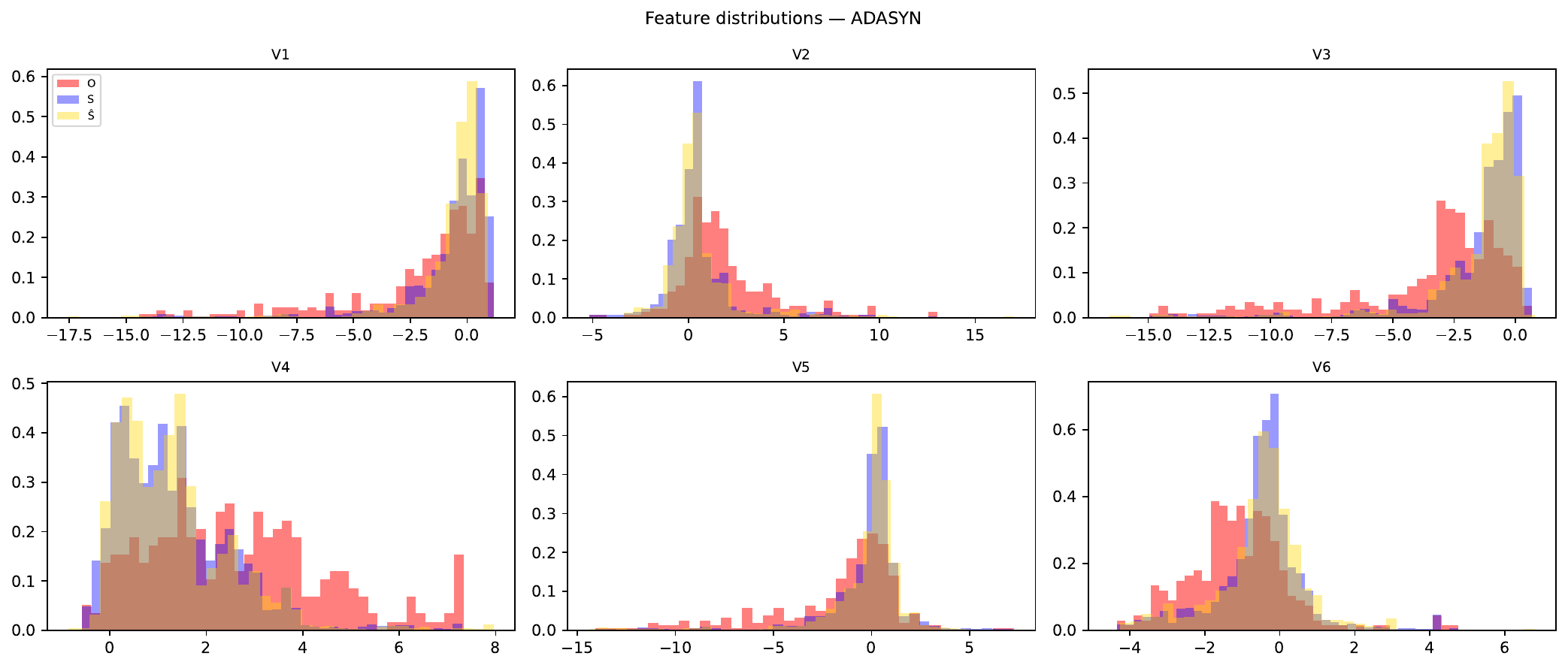}
    \caption{Marginal distributions of six representative features from the Credit Card dataset under ADASYN augmentation.}
    \label{fig:feat_dist}
\end{figure*}

To complement the neighborhood-level view provided by t-SNE, Figure~\ref{fig:feat_dist} examines the effect of the post-processing step directly on the marginal distribution of each feature. 
Figure~\ref{fig:feat_dist} compares the marginal distributions of six representative features from the Credit Card dataset, generated using ADASYN. The raw synthetic data $S$ already captures the general shape of the original distribution $O$ for most features. After post-processing, $\hat{S}$ remains close to $S$ across all six features, indicating that the orthogonal transformation largely preserves the distributional character of the synthetic data while enforcing the correlation constraint for the investigated examples. This is a non-trivial result: despite the heavy constraint imposed on the inter-feature structure, the individual marginal distributions are not significantly distorted. This behavior is consistently observed across all features in the dataset, with the different synthetic data generators. 

\subsection{Downstream Classification Task}\label{sec:classification}

As a supplementary evaluation, we assess the impact of the proposed post-processing step on downstream classification performance in imbalanced settings, with the aim of determining whether it preserves predictive accuracy. We emphasise that correlation matching is not proposed as a classification enhancement technique per se, but rather as a structural correction to synthetic data; any classification benefit is a byproduct of improved data fidelity. Three classifiers: KNN, Gradient Boosting, and a two-hidden-layer MLP are trained on both the raw and Procrustes-corrected augmented sets, then evaluated on a held-out subset of the original data using the F1-score as the primary metric. Each configuration is run five times with different random seeds and the reported scores are averaged. Table~\ref{tab:classif} reports results on Credit Card Fraud ($0.17\%$ minority) and MAGIC Gamma Telescope ($35\%$ minority), across five generation methods.

\begin{table*}[htbp]
\centering
\caption{Classification F1-score (\%) before (B) and after (A) per-class Procrustes correction, with $\Delta = \text{A} - \text{B}$.}
\label{tab:classif}
\setlength{\tabcolsep}{4pt}
\renewcommand{\arraystretch}{1.1}
\begin{tabular}{llccccccccc}
\toprule
 & & \multicolumn{3}{c}{\textbf{KNN}} & \multicolumn{3}{c}{\textbf{GradBoost}} & \multicolumn{3}{c}{\textbf{MLP}} \\
\cmidrule(lr){3-5}\cmidrule(lr){6-8}\cmidrule(lr){9-11}
\textbf{Dataset} & \textbf{Method} & B & A & $\Delta$ & B & A & $\Delta$ & B & A & $\Delta$ \\
\midrule
\multirow{5}{*}{Credit Card}
 & CTGAN          & 84.1 & 83.8 & $-0.30$ & 81.2 & 80.1 & $-1.10$ & 85.3 & 87.2 & $+1.90$ \\
 & SMOTE          & 83.6 & 85.5 & $+1.90$ & 80.4 & 82.2 & $+1.80$ & 85.7 & 85.3 & $-0.40$ \\
 & GaussianCopula & 84.8 & 85.5 & $+0.70$ & 81.0 & 81.4 & $+0.40$ & 85.1 & 86.9 & $+1.80$ \\
 & TVAE           & 83.9 & 84.2 & $+0.30$ & 80.7 & 81.6 & $+0.90$ & 85.4 & 84.6 & $-0.80$ \\
 & ADASYN         & 84.3 & 86.2 & $+1.90$ & 80.9 & 82.8 & $+1.90$ & 85.6 & 86.0 & $+0.40$ \\
\midrule
\multirow{5}{*}{MAGIC}
 & CTGAN          & 73.8 & 74.1 & $+0.30$ & 75.2 & 74.5 & $-0.70$ & 77.4 & 77.9 & $+0.50$ \\
 & SMOTE          & 74.1 & 73.5 & $-0.60$ & 75.6 & 77.4 & $+1.80$ & 77.8 & 78.3 & $+0.50$ \\
 & GaussianCopula & 73.5 & 73.8 & $+0.30$ & 74.9 & 74.1 & $-0.80$ & 77.1 & 77.8 & $+0.70$ \\
 & TVAE           & 74.0 & 74.8 & $+0.80$ & 75.1 & 77.0 & $+1.90$ & 77.5 & 76.8 & $-0.70$ \\
 & ADASYN         & 73.7 & 74.0 & $+0.30$ & 75.3 & 76.2 & $+0.90$ & 77.3 & 77.6 & $+0.30$ \\
\bottomrule
\end{tabular}
\end{table*}

The correction yields on average positive or neutral effects across all classifier--dataset combinations. Average F1-scores are maintained after correction across all classifier--dataset combinations, with only minor variations observed between the pre- and post-correction settings. This is particularly important for the Credit Card dataset, where the scarcity of minority-class observations makes classification performance highly dependent on synthetic data quality. Similar behavior is observed on the MAGIC dataset, suggesting that the correction preserves predictive performance while improving the statistical consistency of the generated data. 
The application of the same protocol across the remaining datasets yields similar results, preserving classification accuracy but providing no measurable performance gain.
Overall, the results show that the proposed post-processing step enhances correlation fidelity without sacrificing classification effectiveness.

\section{Practical Considerations}\label{perlim}

\subsection{Computational Complexity}
In terms of performance, the dominant computational step in the proposed approach is computing the SVD of $\bar{S}(\bar{O}N)^T$, which in general has complexity $\mathcal{O}(p^3)$.
Since $\bar{S}(\bar{O}N)^T$ has rank at most $m$, the complexity reduces to $\mathcal{O}(p^2m)$ when computing a reduced SVD. However, following Remark~\ref{remcomp}, it can be reduced further to $\mathcal{O}(pm^2)$. We emphasize that the QR decomposition trick described in Remark~\ref{remcomp} provides an additional reduction beyond simply computing the reduced SVD of $\bar{S}(\bar{O}N)^T$.

Thus, for $m\ll p$, which is often the case in synthetic data applications, the proposed approach is computationally light.
In cases where $m$ is large, computationally efficient methods that approximate the SVD could be considered. More generally, any computational advantage available for computing the SVD of matrices of the form $BA^T$, with $A,B\in\mathbb{R}^{p\times m}$ and $m < p$, translates naturally to our setting.

\begin{figure}[ht]
    \centering
    \begin{tikzpicture}
        \node at (0,0) {\includegraphics[width=1.0\linewidth]{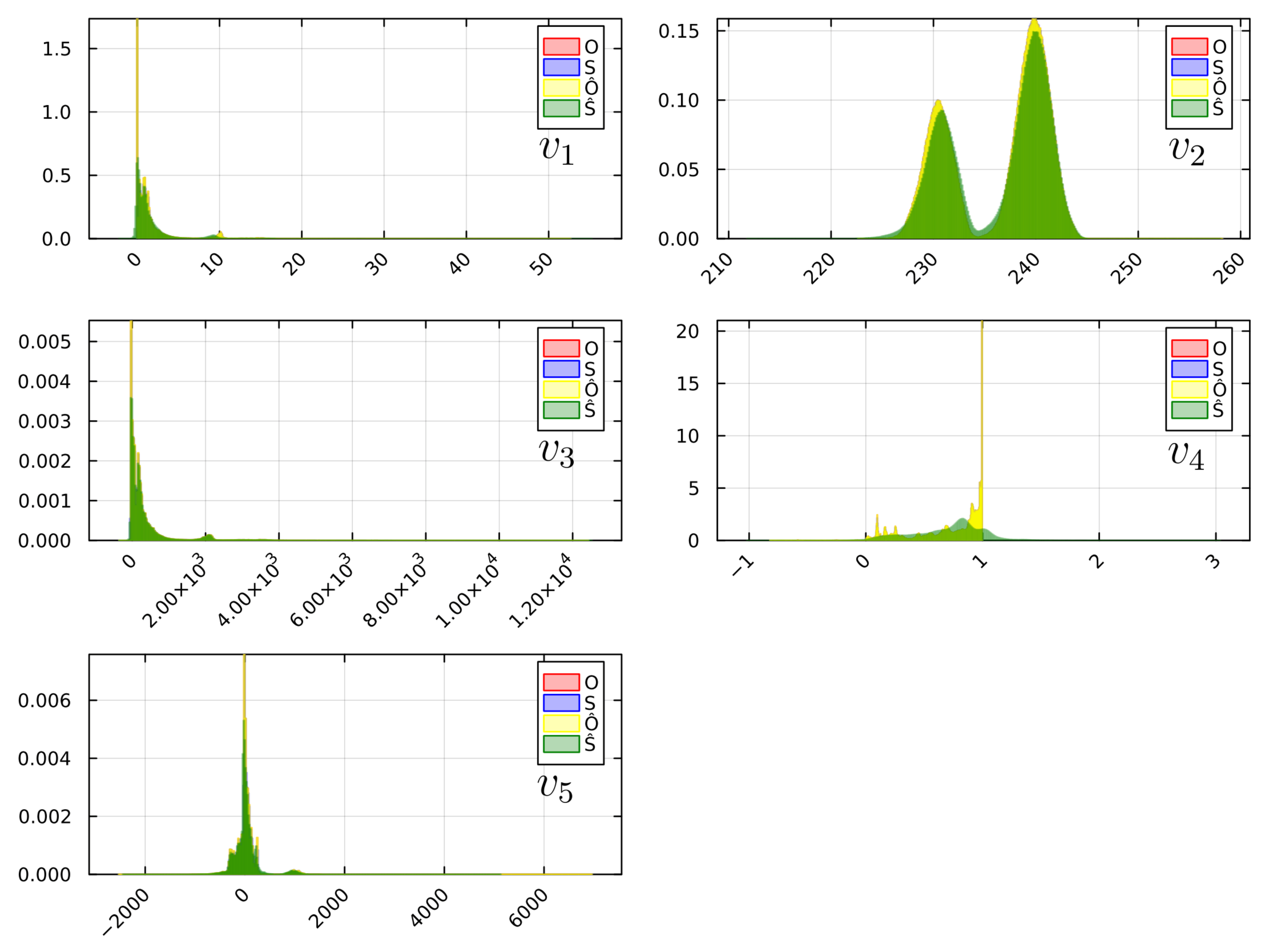}};
    \end{tikzpicture}
    \caption{Distributions of the 5 features in each version of the SustData dataset.}
    \label{fig:SustDist}
\end{figure}

\subsection{Limitations}\label{app}

The distance-minimization nature of our approach inherently gives rise to certain effects worth pointing out. We specifically highlight two side effects to watch out for. First, for features that can only take values inside a specific range, the procedure might create data points outside this range. Second, if the features have different scales, features with smaller scales can be more affected by the transformation, since they have a smaller impact on the least-squares error we aim to minimize.
An illustration is provided by the SustData dataset together with the naive generation approach based on sampling from each individual distribution without regard to any dependence structure. We intentionally consider this setting because it severely distorts the correlation matrix of the original data, making the effect of the proposed correction easier to observe.

Looking at the distributions presented in Figure~\ref{fig:SustDist}, we observe that for features $v_1$ and $v_3$, the procedure produces some negative values that do not make sense for features representing electric power. This is an example of the first effect discussed above, namely that the transformation does not necessarily preserve the admissible range of each feature.
The second effect can also be observed in Figure~\ref{fig:SustDist}. The feature most affected in terms of its distribution is $v_4$. This is due to its small scale compared to the other features, which makes it more susceptible to the least-squares correction.

We emphasize that these two effects are not a bug, but rather a byproduct of the approach working exactly as intended. For instance, the second effect can be accounted for, or even leveraged, through appropriate feature scaling before applying the procedure. The purpose of this discussion is simply to highlight that one should always keep in mind both what the method is designed to do and what may arise as a consequence of that design. Further discussion on how to address these effects is beyond the scope of the present work.

\section{Conclusion}\label{conclusion}

In this study, we treat preserving correlation in synthetic data as a postprocessing problem for an already generated dataset. Our study does not stop at a numerical construction, but gives a full mathematical derivation, with detailed justification of the assumptions in the problem formulation required to guarantee existence and uniqueness of the closed-form solution. We also present a mathematical error estimate and numerically investigate the effects on other relevant properties, such as individual feature distributions, t-SNE geometry, and downstream classification performance, for several datasets and generators. We further discuss computational optimizations and inherent practical implications to inform effective use of the approach.

In future works, the authors are interested in extending this postprocessing viewpoint to other dependence measures, in particular rank-based correlations, and more generally to other statistical properties that may admit similar corrections. Studying the impact of such approaches on additional downstream learning tasks is also of interest.

\ifanonymous
% Identifying declarations and contribution statements are omitted in the double-anonymized review version.
\else
\section*{Conflict of Interest}
The authors declare no conflict of interest.

\section*{Funding}
N.J. and A.M. are supported by the Swedish Energy Agency (SOLVE, grant 52693-1). A.M. is partially supported by the Knowledge Foundation (KK 20200152). O.O. is supported by the MEXT Scholarship.

\section*{CRediT authorship contribution statement}
Oussama Ounissi: Conceptualization, Methodology, Formal analysis, Software, Writing – original draft, Writing – review and editing. Nicklas J\"averg\r{a}rd: Conceptualization, Software, Writing original draft, Writing - review and editing. Assaad Zeghina: Conceptualization, Software, Writing – original draft, Writing - Review and Editing, Visualization. Adrian Muntean: Conceptualization, Funding.

\fi

%% Bibliography
\bibliographystyle{unsrtnat}
\bibliography{ref}



\end{document}